# State of the Art Study of the Safety Argumentation Frameworks for Automated Driving System Safety.

Ilona Cieslik[1][0000-0002-0511-4256]✉, Víctor J. Expósito Jiménez[1][0000−0001−5350−9458], Helmut Martin[1], Heiko Scharke[2], Hannes Schneider[2]

[1] Virtual Vehicle Research GmbH, Graz 8010, Austria
ilona.cieslik@v2c2.at
[2] AVL List GmbH, Graz 8020, Austria

**Abstract.** The automotive industry is experiencing a transition from assisted to highly automated driving. New concepts for validation of Automated Driving System (ADS) include amongst other a shift from a „technology based" approach to a „scenario based" assessment. The safety validation and type approval process of ADS are seen as the biggest challenges for the automotive industry today. Having in mind a variety of existing white papers, standardization activities and regulatory approaches, manufactures still struggle with selecting the best practices that keep aligned with their Safety Management System and Safety Culture. A step forward would be to implement a harmonized global safety assurance scheme that is compliant with relevant regulations, laws, standards, and reflects local rules.

Today many communities (regulatory bodies, local authorities, industrial stakeholders) work on proof-of-concept framework for the Safety Argumentation as an answer to this problem. Unfortunately, there is still no consensus on one definitive methodology and a set of safety metrics to measure ADS safety. An objective of this summary report is to facilitate a comprehensive review and analysis of the literature concerning available methods and approaches for vehicle safety, engineering frameworks, processes of scenario-based evaluation and a vendor- and technology-neutral Safety Argumentation approaches and tools.

**Keywords:** Safety metrics, Safety Argumentation, Automated Driving System, statistical approaches, SOTIF

## 1 An Introduction - The Need for a Safety Argumentation Frameworks

The automotive industry is experiencing a transition from assisted to highly automated driving. The regulatory Framework Document on Automated/Autonomous Vehicles under the United Nations Economic Commission of Europe (UNECE) WP.29 underlines that the certification process of the Automated Driving System (ADS) should prove that „a vehicle shall not cause any non-tolerable risk" [1]. Therefore, one of the biggest challenges to be solved before the release of ADS (i.e., Level 3 and higher) is their safety validation. One of the research questions is: „How can we argue the absence of unreasonable risk of ADS in its defined Operational Design domain (ODD)", without knowing an exact interpretation of „reasonable foreseeable", „preventable" or „tolerable" accidents?

The concepts of „WHAT" to audit and assess when introducing ADS on the public roads reached mature consensus [2], [3], compering to the proof-of-concept frameworks for a Safety Argumentation (i.e., provisions on „HOW" to rational that the ADS is safe enough?). Even though many industrial initiatives and research groups are active in this field, the latter still opens room for stakeholders' discussions. Currently there is no agreement on one definitive methodology with qualitative and quantitative metrics to measure safety, related risks, and social acceptance.

The automotive community agrees that the introduction of ADS requires new assessment and test methods. New approaches for certification and standardization are arising to cover the growing needs of such complex systems. New concepts for validation of ADS include amongst other a shift from a „technology based" approach to a „scenario based" assessment that would allow various types of interventions. Scenario-based testing is already a known and established test approach within the industry for development, certification, and rating purposes [4] but needs to be adopted to the ADS validation purposes to foster public trust in this disruptive technology.

Finally, a step forward would be to implement a safety assurance framework that is embedded in the Safety Management System and aligned with the Safety Culture of the manufacturer. It will help not only to organize safety verification and validation (V&V) in an efficient, objective, transparent, and scalable manner but ensure dense coverage of traffic situations. Additionally, the manufacturer should demonstrate the compliance with relevant regulations, laws, standards, and good practices [5].

The project MASA (Methodology for KPI based ADAS/AD Safety Assessment and Argumentation) is a 2-year long project that started in 2021. Its main objective is to better understand and enhance existing methodologies and numerical approaches for V&V Safety Argumentation frameworks of ADS. The chapter 1 of this publications provides an overview on the MASA project motivation and introduction into ADS V&V needs. The safety and engineering frameworks for ADS considering the regulatory regime, stakeholders' know-how and market demand are summarised in the chapter 2. In the chapter 3 attention is given to the elaboration of the Safety Case approach which is considered by many stakeholders at the center of the Safety Argumentation. The chapter 4 covers the overview on state-of-the-art safety metrics proposed by i) the Contracting Parties to the UNECE and ii) research and industrial projects on novel performance models and proximity metrics. Chapter 5 summarises recommendations on existing Safety Argumentation frameworks and next steps for MASA project, taking into consideration the latest processes prescribed by the ISO 21448 the Safety of The Intended Functionality (SOTIF) [6] community on the risk/hazards and triggering conditions analysis.

## 2 A New Safety Assurance Scheme and Existing Global Initiatives

In the literature there are defined different qualitative and quantitative ways to categorize ADS safety: i) safety as a process/framework, ii) safety as a measurement, and iii) safety as a threshold [7]. In the first stage, MASA focused on understanding assurance frameworks and good practices that consist of interacting methods and tools.

The Safety Argumentation supports the process of gaining knowledge and trust on the operating conditions of ADS. Its main objective is to deliver in a structure way the right evidence to prove safety of the vehicle. Decomposition of the safety requirements prescribed by the regulators and standardization bodies (see chapter 2.1 and 2.2) and implementation of methods for deriving test requirements (see chapter 2.3) help to arrive to credible evidence. A well-structured framework could accelerate the exchange of information and requirements between stakeholders and optimize the tests amount to a manageable minimum [8].

The current automotive and transport communities that dialogue on the enablers for an industrialization of ADS could be split into i) regulatory Informal Working Groups (IWG), ii) standardization activities, and iii) research and industry initiatives.

### 2.1 Regulatory Informal Working Groups for ADS Safety Topics

In 2019 a new Working Party on Automated/Autonomous and Connected Vehicles (GRVA) of the UNECE was created. The WP was tasked with defining provisions for a „safe vehicle" and its approval. Its two IWGs play a significant role in the ADS Safety Argumentation, namely Functional Requirements for Automated and Autonomous Vehicles (FRAV) and Validation Method for Automated Driving (VMAD).

The starting point of the FRAV discussion was a „guardrails approach", where the regulatory bodies decided to not prescribing driving manoeuvres or values. The stakeholder recognise that more than one behaviour could be safe and do not want to hamper the technology development at so early stage of regulatory process. As an outcome of the FRAV sessions a robust list of „General Safety Requirements" with specific criteria was concluded and reported to the GRVA committee [9]. The novelty of this list is that it does not focus on each technology and features separately but addresses the diversity of ADS applications. Moreover, it reflects a global and consolidated view provided by the Member States and the industry representatives covering the full range of anticipated safety needs. Nowadays, the FRAV members supersede where appropriate the list with technical specifications (limits, criteria, formulas, etc.).

On the other hand, the VMAD tackled the certification and audit/assessment process of ADS. The outcome of the discussion is the master document „New Assessment and Test Methods" (NATM) which describes existing test platforms and their interactions. Moreover, the VMAD works on the documentation requirements needed during the audit and assessment phase. The goal is to demonstrate a combination of validation, engineering rigor, post-deployment feedback, and safety culture [4].

## 2.2    Global Standardization Activities

There are many standards that support the Safety Argumentation in different product development phases, from the specification, design, integration, to verification, validation, product release and monitoring. The purpose of those is twofold: i) providing the terms, definition, and summarising existing state of the art of scenario-based validation, ii) defining processes, tools, and check lists. It is relevant to understand their scope and place in the Safety Argumentation framework.

The following standards addressing existing ADS-equipped vehicle safety approaches were considered in MASA roadmap (Tab.1). Their content is intended to be applied to ADS Level 3 and higher according to ISO/SAE 22736 [10].

**Table 1.** Global standardization activities relevant for the Safety Argumentation frameworks.

| Name of the standard: year of publication | Relevance for the ADS Safety Argumentation | Goal: Terms & Definitions (TD) / Process & methods (PM), Check list (CL) |
|---|---|---|
| ISO 3450x series inc. ISO 34501-34505 (u. development) | Terminology and definition of test scenarios for ADS, the overall scenario-based safety evaluation process, a hierarchical taxonomy for ODD, tags for scenarios categorization. | TD, PM, CL |
| ISO/SAE PAS 22736:2021 Taxonomy and definitions for terms related to driving automation systems for on-road motor vehicles | Definitions, taxonomies, and best practices for six levels of driving automation, inc. terms for the dynamic driving task (DDT), DDT fallback, minimal risk conditions, etc. (aligned with SAE J 3016-2021). | TD |
| ISO/PAS 21448:2019 Safety of the intended functionality (SOTIF), rev. in 2022 | Limitations and shortcomings of the technology and the misuse of the function, iterative process of improving the acceptance criteria by triggering conditions and risk/hazards evaluation. | TD, PM |
| ISO 26262:2011 Functional safety (12 parts) | E/E faults and failures as a source of safety problems and calls for a safety case. | TD, PM |
| ISO/AWI TS 5083 Safety for ADS - Design, verification and validation (u. development) | Safety by design, V&V methods based on safety goals and principles, safety case, def. of positive risk balance and avoidance of unreasonable risk (proceed from an ISO/TR 4804) | PM |
| SAE J3131:2022 Automated Driving Reference Architecture | ADS reference software architecture that contains functional modules (does not dictate configuration, neither design requirements) | TD |
| ASAM OpenX standards (concept/implementation) | Foundation of definitions, language, formats, and interfaces for V&V in simulation platforms | TD, PM |
| IEEE P2846:2022 Formal Model for Safety Considerations in Automated Vehicle Decision Making | Set of assumptions and foreseeable scenarios for the development of safety-related models, an extension of the safety envelope violation concept (RSS) | PM |
| ANSI/UL 4600: 2022 Standard for Safety for the | Safety case approach goal-based and technology-agnostic, a catalogue of good practices in the | CL |

| Evaluation of Autonomous Products | whole ADS cycle, inc. safety metrics strategy |
|---|---|

## 2.3 Research and Industry Communities, Initiatives, and Projects

The R&D projects, industrial consortia and local initiatives try to discuss varied approaches related to ADS performance. The task of the MASA was to analyse their unique concepts and what new they bring to the Safety Argumentation. These initiatives not only contribute with the publications and white papers, besides provide valuable input to the regulatory/standardization landscape (see chapter 2.1 and 2.2):

**V&V Methods (Germany)**, a part of PEGASUS project family[1] supported by The Federal Ministry for Economic Affairs and Climate Action (BMWK)**.** The project develops a method, called criticality analysis, which analyzes the open context of urban traffic. It introduces definitions and concepts that support the Safety Argumentation (criticality analysis, criticality phenomenon, criticality metric) [11].
**The CertiCAV Safety Assurance Framework for CAVs (UK)** executed by the Connected Places Catapult on behalf of the Department for Transport**.** The projects developed a framework that has number of novel ideas like a Highly Automated SuperSystem (HASS), the concept of Deployment Risk Classifications, and a foundation for developing requirements with performance indicators [12].
**The SAKURA Safety Assurance KUdos for Reliable Autonomous Vehicles (Japan)** funded by the Ministry of Economy, Trade and Industry (METI)**.** The project proposed a very robust framework that goes further than the nominal driving conditions. It elaborates on safety-relevant disturbances that an ADS may face in a real traffic (traffic, perception, and vehicle motion disturbance scenarios) [13].
**AURORA/UBER Argumentation Concept (USA)**, a commercial initiative. Till now the stakeholders were reluctant to openly share safety cases and reveal the format. Aurora shared their version of the self-driving Safety Case Framework [14].

## 3 Safety Case as a Core Approach to Argue the Safety of ADS

A safety case approach is not a new tool. There are available manuals and standards (e.g., IEC 61508, U.K. Defence Standard 00-55, HSE Railway Safety Case Regulations) which center on a safety case. The good practices from aeronautics, railway, marine or healthcare could have a value in the automotive sector [15]. Literature from R&D projects (see chapter 2.3), University of York community [16] and Edge Case Research studies [17] bring additional input towards the safety case and its format.

---

[1] PEGASUS, Set Level 4to5, V&V Methods are projects developed by the VDA Leitinitiative autonomous and connected driving, https://www.vvm-projekt.de/en/

### 3.1     Safety Case Content and Format

The safety case could be seen as a starting point, as an easy and understandable tool to support the Safety Argumentation. ISO 26262 defines a safety case as an "argument that the safety requirements for an item are complete and satisfied by evidence compiled from work products of the safety activities during development" [18].

The elements of the safety case are i) goal, ii) argumentation, and iii) evidence. When it comes to the argument, it may be deterministic, probabilistic, or qualitative; the evidence may be design, process, or historic experience (with a proof of completeness for coverage of test cases). In the Safety Argumentation it is essential to decompose each goal (claim) further into tangible and measurable performance indicators (results and records). (Safety) metrics are a measurement used to evaluate and track safety performance [19] (see chapter 4.1). Finally, as an outcome of V&V activities on different test platforms, relevant amount of supporting data should be collected to produce quantitative evidence of safety case credibility.

A format for structuring argumentation should be notation agnostic (i.e., textual, tabular, graphical). Some manuals like widely used GSN (The Goal Structuring Notation [16]) or MISRA Safety Case Guideline [20] support the arguments creation and documentation process. Using an agreed format enables a consistent and traceable decomposition from claims down to V&V methods and keeping the safety case evolving over the life cycle (a living document).

### 3.2     Acceptance Criteria

Acceptance criteria for ADS are based on a statement that „safe enough" is not just a number, it is an argument. In literature there are many strategies and accompanying argumentation patterns, e.g., safety case architecture as in UL 4600 [21] or PAS 1881:2022 [22] include references to ALARP (Reducing risk as low as reasonably practicable). Some of acceptance criteria widely used in other sectors are summarized in [23], those are: GAMAB („generally at least as good as"), MEM (Minimum Endogenous Mortality), MISRI (Minimum Industry Safety Return on Investment) or RAPEX (Rapid Exchange of Information), amongst other.

Currently more attention gain „a positive risk balance" criteria. In this context, ADS should generate a „statistical positive risk balance" such that ADS demonstrate superior performance when compared statistically against human driving performance. One of the objectives of a new ISO 5083 standard under development [24] is to structure the holistic safety approach with the safety case pattern for ADS based on recommendations of the German Ethics Commission in 06/2017 (BMVI) [25].

## 4      Safety Evidence: An Attempt to „Quantify" the ADS Safety

As explained in the introduction, the scenarios are one of the main factors in the V&V strategy. They are one approach to link testing activities to an argumentation strategy. There are different ways of classifying the scenarios, according to i) the abstraction and detail level of parameters - functional, logical, and concrete scenarios [26] ii) probability of occurrence in the real-world traffic vs. risk of the scenario - typical, critical, edge case scenario [27], iii) the new draft regulation for the type-approval of the ADS by the EC – nominal, critical, failure scenarios [3].

The current data bases of crashes and statistics often provide basic data on a high and aggregated level, without diving into the cause of the crash. For ADS there should be different types of the safety goals that are included in the argumentation process [28] and that support accident investigation and decomposition process:

**Functional Safety Goals -** according to ISO 26262 compliance - behavior in case of system or component failure (hazards) that may arise by the functionality in the E/E system within its ODD are identified and assigned to the required ASIL level [18],
**SOTIF Goals -** according to ISO 21448 compliance - triggering conditions (including foreseeable misuses) and obtained test results purpose is to reduce the residual risk to an acceptable level and to improve ADS nominal performance [6],
**Ethics Goals and Societal Expectations** - according to ethical standards (e.g., ISO 39003 u. development), human errors and intentions, that could endanger other actors lives or damage properties, discriminate road users by age, gender, clothes, handicap,
**Laws and Regulations Goals** - according to UNECE regulations, national and traffic law under Road Traffic Code and locally accepted behavioural laws respectively,
**Cybersecurity** - according to ISO/SAE 21434, SAE J3061 (not part of this paper).

The MASA project focuses mainly on prescriptive requirements coming from laws, regulations, and standards (see chapter 4.1), and investigation of novel approaches to support SOTIF processes (see chapter 4.2).

### 4.1 Requirements-Based Testing With Numerical Approaches

The chapter 2 explains the need to quantify the product safety, its social acceptance and liability. Indeed, the introduction of ADS requires new type of performance models and metrics. Consolidation of safety related metrics was the main objective of MASA, for ADS none-safety related metrics (network efficiency, energy emission, drive quality, costs and public health or comfort), the reader could consult [29].

**Performance Models** are based on „roadmanship" concept, which means the ability to drive on the road safely without creating hazards and responding correctly to hazards created by others. They reflect the ADS's situational awareness, time to response, speed adjustment, the vehicle's physics, driving culture and laws and diverse driving scenarios [7]. In the literature, ADS performance models are named as mathematical models for trajectory planning, safety envelope or escape path. Some prominent models that are part of the ADS regulations and standards are:

*Competent and Careful Human Driver's Performance Model (C&C)* proposed by the Japanese delegation to the UNECE. The model is included in the UN Regulation No. 157 - Automated Lane Keeping Systems (ALKS), Annex 3. Its main assumption is that traffic accidents are split into rationally foreseeable and preventable [30].

*The Responsibility-Sensitive Safety Model (RSS)* by Intel (US). It is a white-box mathematical model. It formalizes the „duty of care", which means that an individual should exercise „reasonable care" while performing acts that could harm others [31]. The safety envelope concepts are reflected in IEEE 2846 standard [32].
*Fuzzy Safety Model (FSM)* of the Joint Research Centre of the European Commission. This model builds on RSS findings. Its characteristics are based on fuzzy logic that would not require the vehicle to decelerate very sharply or very often. The model is considered next to the C&C model in the draft of ALKS extension (Annex 3) under the leadership of SIG UNR157 Task Force of UNECE [33].

The reader should be aware of other models existing in the literature, that were analysed under MASA but currently are rather in the concept phase: Safety Force Field (SFF), Instantaneous Safety Metric (ISM), Criticality Metric using Model Predictive Trajectory Optimization [19].

**Safety Metrics and Their Thresholds** - till today the universal metrics for road safety were historical crash data like frequency and severity. The existing test procedures and protocols for assisted functions (ADAS) like Euro NCAP consumer tests or regulatory documents (UN No. 79, UN No. 131) introduce simple but comprehensive metrics to select situations out of traffic events to reduce the amount of test effort.

The need for harmonization and standardization of terms, definitions, and techniques for ADS safety measures has been recognized for last decade [34]. The metrics that refer to testing in nominal conditions are used under exchangeable terminologies. They evaluate the criticality of the traffic situation: Behavioral Safety Measures [35], Safety Performance Assessment Metrics [36], Proximal Surrogate Indicators, Temporal/Spatial-based Conflict Indicators [37], Criticality/Risk Metrics [11].

Mentioned above references define important properties of each metric: definition, taxonomy, data source from off or onboard sources, observable variables, formulation: mathematical model, assumptions/thresholds when applicable, origin, limitations and advantages, reason for inclusion, research examples, type of scenario, manoeuvre collision type suitability. The section „16. Metrics and Safety Performance Indicators (SPIs)" of [21] gives additional guidelines on metrics as a part of the Safety Culture. The manufacturer should present a metric strategy with collection, evaluation, and improvement processes. Further, the safety metrics could be categorized into [7]:

*Prior/Predictive (Leading Metrics)* - including general performance characteristics, associated with vehicle kinematics (longitudinal and lateral distance), perception and assessment of Object and Event detection specification (OEDR) [38], safe motion control metrics etc. Leading metrics are particularly important for ADS because their events happen more frequently than lagging measures.
*An Outcome (Lagging Metrics)* - covering post deployment, longer term metrics like for driver disengagement [39], ODD metrics [40], violation of road rules and crash severity and frequency. Assessing a correlation of leading metrics to safety outcomes should be used to drive improvement of the metrics and thresholds [19].

The metrics could only be implemented successfully considering the threshold and the pass-fail criteria definition. Their definition is not an easy task, neither standardized (e.g., ADAS systems use fixed rules of thumb like the two-second rule for establishing safety envelope) [41]. Additionally, [21] states that the thresholds could be a desired value, limit, or incident frequency. The authors recommend varied approaches for consolidating claims of different stakeholders when selecting values/targets: i) technology aspects (the state-of-the-art technology limitations), ii) human driver aspects (response to traffic events: perception, recognition, decision), iii) social aspects (socially acceptable behaviours) iv) legal aspects (decisions of the court jurisdiction). Moreover, to allow ADS deployment, the thresholds should support federal, state, and local laws and could be a function of several parameters such as vehicle capabilities, road user type, and speed of the ego vehicle [19].

### 4.2 Quantitative Hazard and Risk Analysis as a Part of the SOTIF Processes

The safety metrics and performance models (see chapter 4.1) could support argumentation for known and nominal conditions. When it comes to edge and rare traffic scenarios, the probabilistic methods with the use of virtual testing environment could play a crucial role. A challenge for ADS today is a sound and systematic methodology for the identification and quantification of scenarios that are likely to exhibit hazardous behaviour.

SOTIF focuses on the limitations of the used technology as well as the misuse of the function. One of its work product is to discover the „potential triggering events" with the purpose of improving the defined acceptance criteria and minimizing the known/unknown hazard scenario with each new iteration [6]. The MASA authors identified exemplary triggering conditions (TC) and categorized them into 5 groups: i) Environmental TC (weather state, illumination, impact on the quality of the road surface), ii) Infrastructural TC (geometry of the road, road furniture, objects on the road/surroundings), iii) Communication and other interferences, iv) Other road actors (adverse traffic behaviour, non-standard road participants, surrounding vehicles), v) Ego vehicle behaviour (perception (OEDR), operation/ maneuverers (DDT)). What is novel in MASA, selected triggering events will be parametrized and demonstrated in industrial settings like V&V Tool Chain of AVL partner.

Parametrization process of the SOTIF triggering conditions is a robust task as it requires new type of statistics (unit/scale, boundary values, ground truth measurement process, source of potential statistics, etc.). Real world occurrence likelihoods of today's data (traffic and crash data, labelled data sets from the naturalistic driving, weather forecast records, or infrastructure maintenance reports) require a totally new way of looking at them. The triggering conditions dependencies and dependencies between triggering events and the scenarios constitute another challenge [42].

Currently, the authors investigate and compare preliminary statistical approaches for SOTIF safety argumentation, and the outcome will be reported in the second year. Some of the prominent approaches taken into consideration include [35], [42], [43]. The main objective of the MASA project in the next phase is to enhance the existing workflow of the SOTIF standard with novel numerical approaches.

## 5      Conclusions and Outlook

The MASA project examined concepts for safety metrics, formal models, taxonomies, and process approaches for measuring the safety of ADS-equipped vehicles. The analysis of the available safety frameworks revealed the current research needs and existing gaps between regulatory and industrial pace. Today's regulatory documents around ADS V&V type approval leave a lot open to interpretation.

Determining metrics may help to demonstrate safety as a part of holistic approach for assessing/evaluating aspects of ADS safety. Unfortunately, no one has presented a fully suitable set of metrics for arguing safety of ADS across the range of its functions and features, use cases, and ODDs. The manufacturer could follow many available safety practices to decide which fit the best in their Safety Case, but still depend on the Type Approval Authority opinion of whether it complies with requirements.

On the other hand, the global consensus was reached on the need to develop ADS strategy for safety metrics. It is seen as a joint work of stakeholders from industry, academia, authorities, and consultation with the civil society. The members of the working groups and research projects recognize both quantitative (product-oriented test results and records) and qualitative approaches (the Safety Culture and Management System oriented on processes) as valid to understand the level of safety and duty of care. It still needs to be explored how to leverage a mix of those approaches.

Further, it is suggested that for level 3 ADS-equipped vehicles, the current practices coming from the "technology based" regulatory and standardization documents should be extended with novel standards, such as ISO 26262, ISO/SAE 21448 (and cybersecurity). In this report, we only give an overview of the state of the art around statistical approaches for hazard and risk analysis to identify rare and unknown cases.

The goal of the next period of MASA project is to combine and extend established techniques for hazard analysis and risk assessment. The challenge is to supersede traffic scenario databases and test case frameworks with low probability but high consequence events (in literature so called Triggering Conditions [6], disturbance scenarios [13], or criticality phenomena [11]) that are not captured in the existing database of global functional scenarios [4]. Unfortunately, there is still lack of publicly acceptable and systematic identification method for triggering events.

To conclude, the development of a reliable safety measure will be a significant achievement that expands the current V&V methods. Moreover, extending nominal and well-known scenarios with SOTIF approaches could create additional trust in ADS complex technology.

## 6      Acknowledgement

The publication was written at Virtual Vehicle Research GmbH in Graz, Austria. The authors would like to acknowledge the financial support within the COMET K2 Competence Centers for Excellent Technologies from the Austrian Federal Ministry for Climate Action (BMK), the Austrian Federal Ministry for Digital and Economic Affairs (BMDW), the Province of Styria (Dept. 12) and the Styrian Business Promotion Agency (SFG). The Austrian Research Promotion Agency (FFG) has been au-thorised for the programme management. They would furthermore like to express their thanks to their supporting industrial project partner, namely AVL List GmbH.

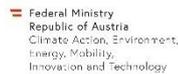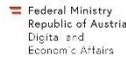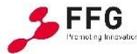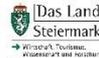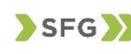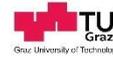